\documentstyle[12pt,worldsci,epsf]{article}
\begin{document}
\title{GLUON EXCITATIONS OF THE STATIC-QUARK POTENTIAL}
\author{K.J.~Juge, J.~Kuti, and 
 C.~Morningstar\thanks{Talk presented by C.~Morningstar}\\
 {\em University of California at San Diego,
  La Jolla, California 92093-0319}}
\maketitle
\setlength{\baselineskip}{2.6ex}

\vspace{0.7cm}
\begin{abstract}

The spectrum of gluon excitations in the presence of a static
quark-antiquark pair is presented.  Our results are obtained from
computer simulations of gluons on anisotropic space-time lattices
using an improved gauge-field action.  Measurements for quark-antiquark
separations $r$ ranging from 0.1 fm to 4 fm and for various orientations
on the lattice are made.  Discretization errors and finite volume 
effects are taken into account.  Surprisingly, the spectrum does not
exhibit the expected onset of the universal $\pi/r$ Goldstone excitations
of the effective QCD string, even for $r$ as large as 4 fm.

\end{abstract}
\vspace{0.7cm}

\section{Introduction}

Accurate knowledge of the properties of the stationary states of
glue in the presence of the simplest of color sources, that of a static
quark and antiquark separated by some distance $r$, is an
important stepping stone on the way to understanding confinement.
It is generally believed that at large $r$, the linearly-growing
ground-state energy of the glue is the manifestation of the confining
flux whose fluctuations can be described in terms of an effective
string theory.  The lowest-lying excitations are then
the Goldstone modes associated with spontaneously-broken transverse
translational symmetry.  Expectations are less clear for small $r$.
The determination of the energies of glue in the presence of a static
quark-antiquark pair is also the first step in the Born-Oppenheimer
treatment of conventional and hybrid heavy-quark mesons\cite{hasenfratz}.
A better understanding of hybrid quarkonium should provide valuable
insight into the nature of light hybrid mesons, which are currently of
great experimental and theoretical interest.

Even the simplest property, the energy spectrum, of the stationary
states of glue interacting with a static quark-antiquark pair is not
accurately known.  The main goal of this work is to remedy this.  Here,
we present, for the first time, a comprehensive determination of the
low-lying spectrum of gluonic excitations in the presence of a static
quark-antiquark pair.  In this initial study, the effects of light
quark-antiquark pair creation are ignored.  A few of the energy levels for
$r$ less than 1 fm have been studied before\cite{previous}.  Our results
for these quantities have significantly improved precision, and we
have extended the range in $r$ to 4 fm.  Most of the energy levels
presented here have never been studied before.  Some of our results
were previously reported\cite{earlier}.

\section{Computation of the energies}

We adopt the standard notation from the physics of diatomic molecules
and use $\Lambda$ to denote the magnitude of the eigenvalue of the projection
$\vec{J_g}\!\cdot\hat{\bf r}$ of the total angular momentum $\vec{J_g}$
of the gluons onto the molecular axis $\hat{\bf r}$. The capital Greek
letters $\Sigma, \Pi, \Delta, \Phi, \dots$ are used to indicate states
with $\Lambda=0,1,2,3,\dots$, respectively.  The combined operations of
charge conjugation and spatial inversion about the midpoint between the
quark and the antiquark is also a symmetry and its eigenvalue is denoted by
$\eta_{CP}$.  States with $\eta_{CP}=1 (-1)$ are denoted
by the subscripts $g$ ($u$).  There is an additional label for the
$\Sigma$ states; $\Sigma$ states which
are even (odd) under a reflection in a plane containing the molecular
axis are denoted by a superscript $+$ $(-)$.  Hence, the low-lying
levels are labelled $\Sigma_g^+$, $\Sigma_g^-$, $\Sigma_u^+$, $\Sigma_u^-$,
$\Pi_g$, $\Pi_u$, $\Delta_g$, $\Delta_u$, and so on.  For convenience,
we use $\Gamma$ to denote these labels in general.

The glue energies $E_\Gamma(\vec{r})$ were extracted from Monte Carlo estimates
of generalized Wilson loops.  The expectation value of the path-ordered
exponential of the gauge field along a closed loop is known as a Wilson
loop.  The familiar static-quark potential $E_{\Sigma_g^+}(r)$ can be
obtained from the large-$t$ behaviour $\exp[-tE_{\Sigma_g^+}(r)]$
of the Wilson loop for a rectangle of spatial length $r$ and temporal
extent $t$.  The two spatial segments of such a Wilson loop can be
replaced by a complicated {\em sum} of spatial paths, all sharing the
same starting and terminating sites, in order to extract the
$E_\Gamma(\vec{r})$ energies, as long as the sum of paths transforms
as $\Gamma$ under all symmetry operations.  Using several different
sums of paths (we typically use 3 to 22 such sums) then produces a
matrix of Wilson loop correlators $W_\Gamma^{ij}(r,t)$.

Monte Carlo estimates of the $W_\Gamma^{ij}(r,t)$ matrices
were obtained in eight simulations using an improved gauge-field
action\cite{peardon}.  Our use of anisotropic lattices in which the temporal
lattice spacing $a_t$ was much smaller than the spatial spacing $a_s$
was crucial for resolving the glue spectrum, particularly for large $r$.
To hasten the onset of asymptotic behaviour, iteratively-smeared spatial
links\cite{peardon} were used in the generalized Wilson loops.
The temporal segments in the Wilson loops were constructed from
thermally-averaged links, whenever possible, to reduce statistical noise.
To facilitate the removal of finite-spacing errors by extrapolating to
the continuum limit, results for several values
of the lattice spacing were obtained; simulations were done using
$a_s\approx 0.29,0.27,0.22,0.19$, and $0.12$ fm.  Two runs corresponding
to the same $a_s$ but different $a_t$ were done to provide a measure
of the $a_t^2$ errors in our results (our action has $O(a_t^2,a_s^4)$ errors).
Such information is important for carrying out the $a_s\!\rightarrow\! 0$
extrapolations.  Agreement of energies obtained using different 
quark-antiquark orientations on the lattice was used to check the smallness of
finite-spacing errors and to help identify the continuum $\Lambda$ value
corresponding to each level (there are only discrete symmetries on the lattice).
For this reason, results were obtained not only for cases in which the
molecular axis coincided with an axis of the lattice, but also for
separation vectors such as $(r,r,r)/\sqrt{3}$ and $(r,r,0)/\sqrt{2}$.

The matrices $W_\Gamma^{ij}(r,t)$ were reduced in the data fitting phase
to single correlators and $2\times 2$ correlation matrices using the
variational method.  The lowest-lying glue energies were then extracted
from these reduced correlators by fitting a single exponential and a sum
of two exponentials, the expected asymptotic forms, in various ranges
$t_{\rm min}$ to $t_{\rm max}$ of the source-sink separation.  The 
two-exponential fits were used to check for consistency with the
single-exponential fits, and in cases of favourable statistics, to extract
the first-excited state energy in a given channel.

Three additional runs on small lattices were done to verify that
finite-volume errors in our results were negligible.  We confirmed the
smallness of the $a_s/a_t$ renormalization for two values of the QCD
coupling by extracting the ground state potential from Wilson
loops in different orientations.  The hadronic scale parameter 
$r_0\approx 0.5$ fm
was used to determine the lattice spacing\cite{peardon}. The additive
ultraviolet-divergent self-energies of the static sources were
removed by expressing all of our results with respect to 
$\Sigma_g^+(r_0)$.  Finite-lattice spacing errors were removed
by extrapolating our simulation results for 
$r_0[E_\Gamma(r)-E_{\Sigma_g^+}(r_0)]$ to the $a_s\!\rightarrow\! 0$
continuum limit.  These extrapolations were carried out by fitting all
of our simulation results to an ansatz 
$F_{\rm cont}(r) + a_s^4\ F_{\rm latt}(r)$: a ratio of a polynomial
of degree $p+1$ over a polynomial of degree $p$, where $p=1$ or $2$,
was found to work well for the continuum limit form $F_{\rm cont}(r)$,
and $F_{\rm latt}(r)$ was chosen empirically to be a sum of three terms
$1/\sqrt{r}$, $1/r$, and $1/r^2$.  All fits yielded $\chi^2/{\rm dof}$
near unity.  Continuum $\Lambda$ values were easily identified in all
cases but one: we were unable to distinguish between a $\Pi_u^\prime$
and $\Phi_u$ interpretation for the on-axis $E_u^\prime$ level.

\section{Results}

\begin{figure}[t]
\begin{center}
\leavevmode
\epsfxsize=2.9in\epsfbox[78 184 532 678]{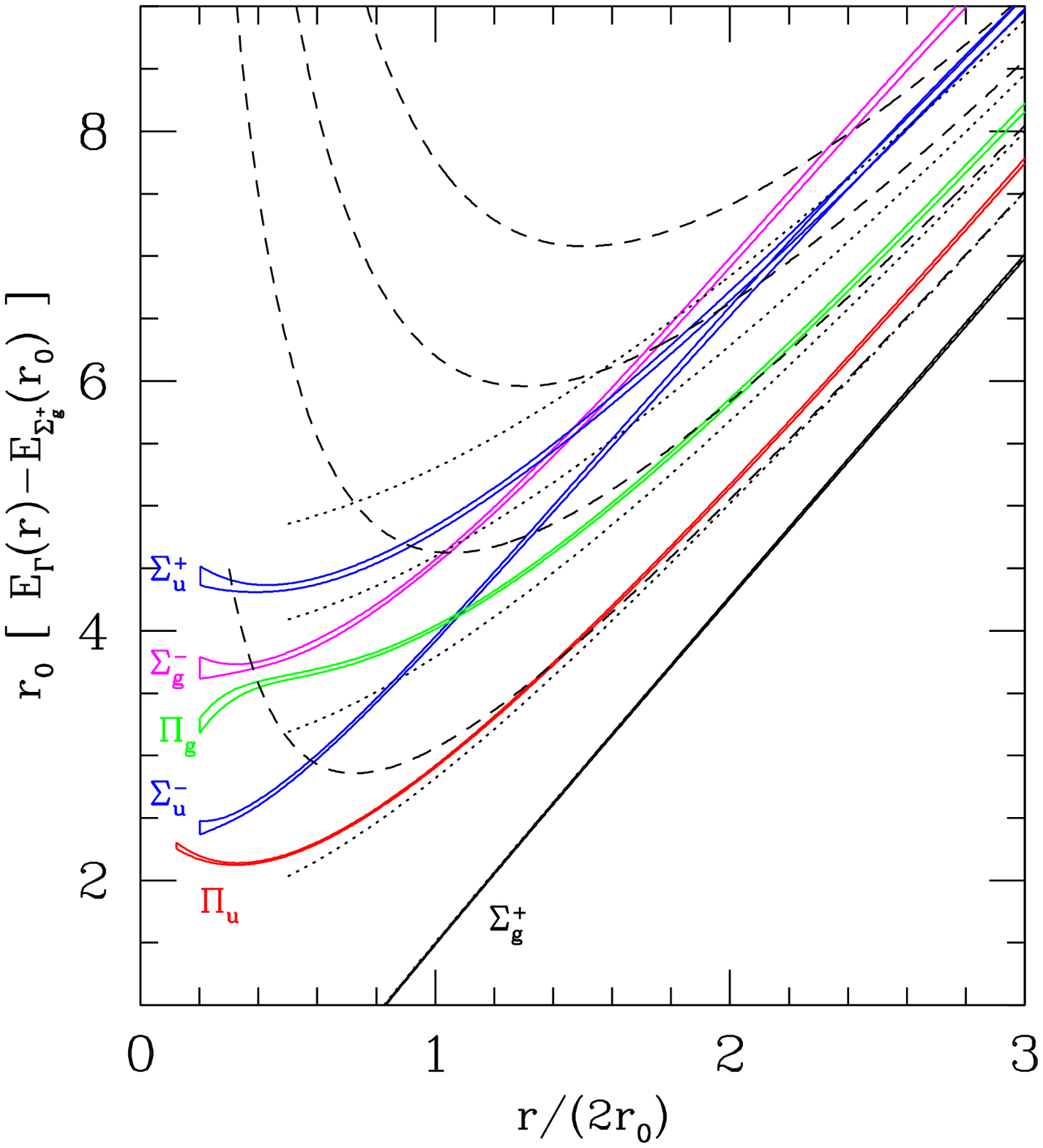}
\epsfxsize=2.9in\epsfbox[78 184 532 678]{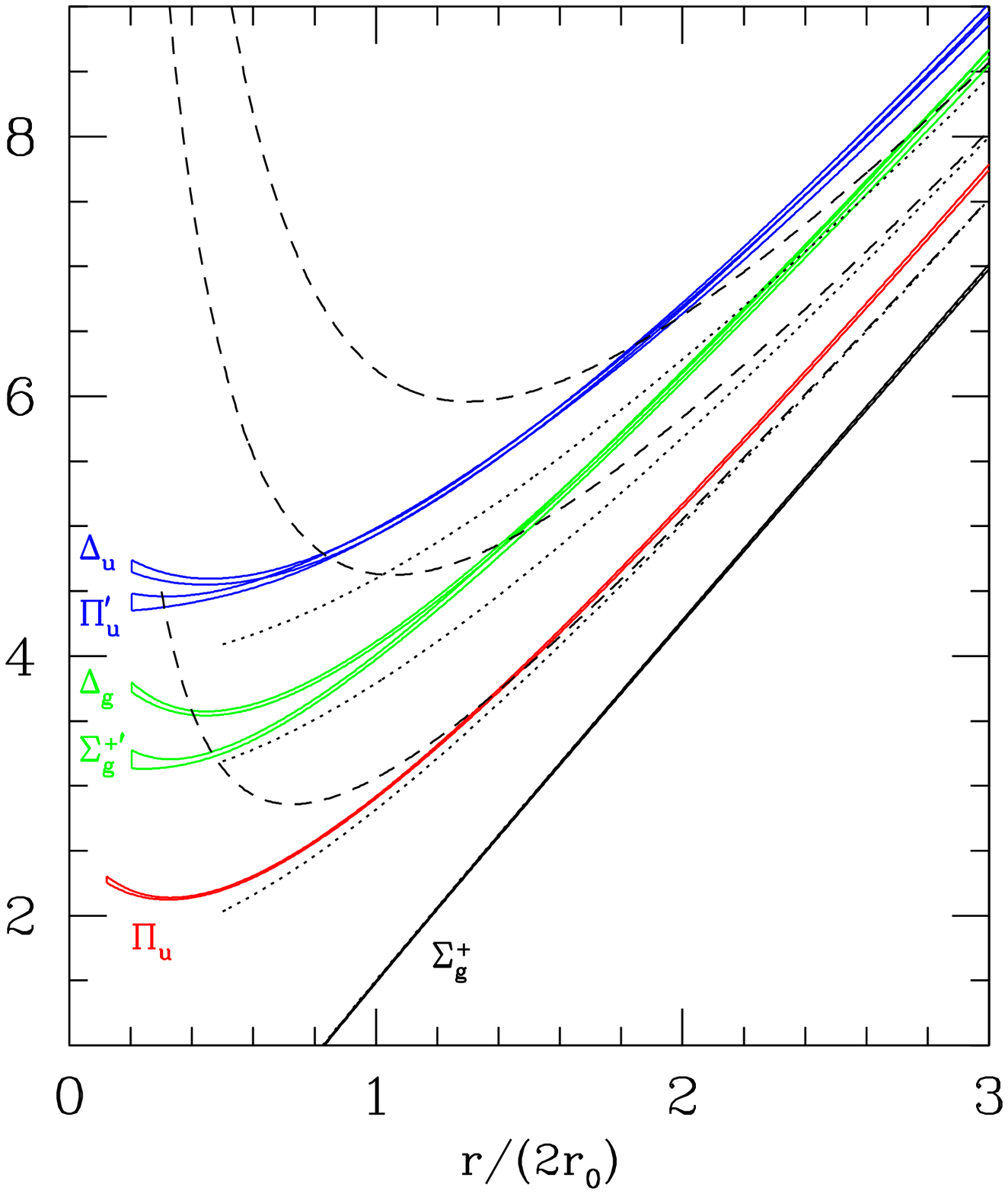}
\end{center}
\caption[figresone]{Plot of the continuum-limit extrapolations (with
  uncertainties as indicated) for
  $r_0 [E_\Gamma(r)-E_{\Sigma_g^+}(r_0)]$ against $r/(2r_0)$ for
  various $\Gamma$.  The dashed lines indicate the locations of
  the $m\pi/r$ gaps above the $\Sigma_g^+$ curve for $m=1$, 2, 3, and 4.
  The dotted curves are the naive Nambu-Goto energies in four-dimensions.
  Note that we cannot rule out a $\Phi_u$ interpretation for the
  curve labelled $\Pi_u^\prime$.
  }
\label{fig:res}
\end{figure}

Our continuum-limit extrapolations are shown in Fig.~\ref{fig:res}.
The ground-state $\Sigma_g^+$ is the familiar static-quark potential.
A linearly-rising behaviour dominates the $\Sigma_g^+$ potential once
$r$ exceeds about 0.5 fm and we find no deviations from the linear form
up to 4 fm.  The lowest-lying excitation is the $\Pi_u$.  There is
definite evidence of a band structure at large $r$: the $\Sigma_g^\prime$,
$\Pi_g$, and $\Delta_g$ form the first band above the $\Pi_u$;
the $\Sigma_u^+$, $\Sigma_u^-$, $\Pi_u^\prime/\Phi_u$, and $\Delta_u$
form another band.  The $\Sigma_g^-$ is the highest level at large $r$.
This band structure breaks down as $r$ decreases below 2 fm. In particular,
two levels, the $\Sigma_g^-$ and $\Sigma_u^-$, drop far below their
large-$r$ partners as $r$ becomes small.  Note that for $r$ above 0.5 fm,
all of the excitations shown are stable with respect to glueball decay.
As $r$ decreases below 0.5 fm, the excited levels eventually become
unstable as their gaps above the ground state $\Sigma_g^+$ exceed the
mass of the lightest glueball.

A universal feature of any low-energy description of a thin fluctuating
flux tube is the presence of Goldstone excitations associated with the 
spontaneously-broken transverse translational symmetry.  These transverse
modes have energy separations above the ground state given
by multiples of $\pi/r$.  The level orderings and approximate degeneracies
of the gluon energies at large $r$ match, without exception, those expected
of the Goldstone modes.  However, the precise $m\pi/r$ gap behaviour is not
observed (see Fig.~\ref{fig:res}).  For separations less than 2 fm, the
gluon energies lie well below the Goldstone energies and the Goldstone
degeneracies are no longer observed.  The two $\Sigma^-$ states are in
violent disagreement with expectations from a fluctuating string.  The
gluon energies also cannot be explained in terms of a Nambu-Goto string
(whose quantization in four dimensions is problematical, besides).

These results are rather surprising.  Our results cast serious doubts
on the validity of treating glue in terms of a fluctuating string
for quark-antiquark separations less than 2 fm.  Note that such a
conclusion does not contradict the fact that the $\Sigma_g^+(r)$ energy
rises linearly for $r$ as small as 0.5 fm.  A linearly-rising term is not
necessarily indicative of a string: for example, the adiabatic bag model
predicts a linearly-rising ground state much before the onset of
string-like behaviour, even in the spherical approximation\cite{hasenfratz}.
For $r$ greater than 2 fm, there are some tantalizing signatures of
Goldstone mode formation, yet significant disagreements still remain.
To what degree these discrepancies can be explained in terms of
a distortion of the Goldstone mode spectrum arising from the spatial
fixation of the quark and antiquark sources is currently under
investigation.  This work was supported by the U.S.~DOE,
Grant No.\ DE-FG03-97ER40546.
\vspace{0.3cm}

\end{document}